\documentclass[aps,prd,superscriptaddress,twocolumn,nofootinbib,amsmath,amssymb,amsfonts,nokeys,nopacs]{revtex4}
\usepackage{dcolumn}      % Align table columns on decimal point
\usepackage{bm}           % bold math
\newcommand{\be}{\begin{equation}}
\newcommand{\ee}{\end{equation}}
\newcommand{\eei}{\end{equation}}
\newcommand{\bc}{\begin{center}}
\newcommand{\ec}{\end{center}}
\newcommand{\ber}{\begin{eqnarray}}
\newcommand{\ear}{\end{eqnarray}}
\newcommand{\ba}{\begin{array}}
\newcommand{\ea}{\end{array}}
\newcommand{\n}{\widetilde{\nabla}}
\newcommand{\na}{\nabla}

\newcommand{\hs}{\,-\,}
\newcommand{\p}{\partial}
\newcommand{\da}{{\cal D}_a}

\newcommand{\inps}{\psi^{-1}}

\newcommand{\ls}{{\cal L}}

\newcommand{\sfrac}[2]{{\textstyle{#1\over#2}}}
\def\case#1/#2{\textstyle\frac{#1}{#2} }

\newcommand{\dd}{\mathrm{d}}
\newcommand{\bk}{\bm{k}}
\newcommand{\bq}{\bm{q}}
\newcommand{\bp}{\bm{p}}
\newcommand{\unitk}{\hat{k}}
\newcommand{\unitbk}{\hat{\bm{k}}}
\newcommand{\bx}{\bm{x}}

\renewcommand{\div}{\hskip0.9pt{\mathsf{div}\hskip2pt}}
\newcommand{\curl}{\hskip0.9pt{\mathsf{curl}\hskip2pt}}
\def\lgl{\langle}
\def\rgl{\rangle}
\renewcommand{\S}{_{\hskip-0.8pt\mathrel{\vcenter{\hbox{\tiny\ooalign
{\raise 1.5pt\hbox{\textsf{S}}}}}}}}
\newcommand{\V}{_{\hskip-0.8pt\mathrel{\vcenter{\hbox{\tiny\ooalign
{\raise 1.5pt\hbox{\textsf{V}}}}}}}}
\newcommand{\T}{_{\hskip-0.8pt\mathrel{\vcenter{\hbox{\tiny\ooalign
{\raise 1.5pt\hbox{\textsf{T}}}}}}}}

\newcommand{\acc}{\dot{u}}

\begin{document}
%%%%%%%%%%%%%%%%%%%%%%%%%%%%%%%%%%
\title{Gravitational waves generated by second order effects during inflation}
%%%%%%%%%%%%%%%%%%%%%%%%%%%%%%%%%%

\author{Bob Osano}
 \affiliation{Cosmology \& Gravity Group, Department of Mathematics and Applied Mathematics,
 University of Cape Town, Rondebosch 7701, South Africa}

\author{ Cyril Pitrou}
 \affiliation{Institut d'Astrophysique de Paris,  Universit\'e Pierre~\&~Marie Curie - Paris VI,
CNRS-UMR 7095, 98 bis, Bd Arago, 75014 Paris, France.}

\author{Peter Dunsby}
\affiliation{Cosmology \& Gravity Group, Department of Mathematics and Applied Mathematics,
  University of Cape Town, Rondebosch 7701, South Africa}
\affiliation{South African Astronomical Observatory, Observatory
7925, Cape Town, South Africa}

\author{Jean-Philippe Uzan}
\affiliation{Institut d'Astrophysique de Paris,  Universit\'e Pierre~\&~Marie Curie - Paris VI,
CNRS-UMR 7095, 98 bis, Bd Arago, 75014 Paris, France.}

\author{Chris Clarkson}
\affiliation{Cosmology \& Gravity Group, Department of Mathematics and Applied Mathematics,
 University of Cape Town, Rondebosch 7701, South Africa}
\affiliation{Institute of Cosmology and Gravitation, University of
Portsmouth, Mercantile House, Portsmouth PO1 2EG, Britain}
%%%%%%%%%%%%%%%%%%%%%%%%%%%%%%%%%%

\date{\today}

\begin{abstract}

The generation of gravitational waves during inflation due to the non-linear coupling of scalar
and tensor modes is discussed. Two methods  describing gravitational wave perturbations are 
used and compared: a covariant and local approach, as well as a metric-based analysis based on the Bardeen formalism. An application to slow-roll inflation is also described.
\end{abstract}

 \maketitle
%%%%%%%%%%%%%%%%%%%%%%%%%%%%%%%%%%%
\section{Introduction}\label{sec1}
%%%%%%%%%%%%%%%%%%%%%%%%%%%%%%%%%%%

The generation of gravitational waves (GW) is a general prediction of an early inflationary
phase~\cite{infgw}. Their amplitude is related to the energy scale of inflation and they are
potentially detectable via observations of B-mode polarization in the cosmic microwave background
(CMB) if the energy scale of inflation is larger than $\sim 3\times10^{15}$~GeV~\cite{MKam,USeljak,MKes,LKnox,USeljakH}. Such a detection would be of primary importance to test inflationary models.

Among the generic predictions of one-field inflation~\cite{inflgen} are the existence of
(adiabatic) scalar and tensor perturbations of quantum origin with an almost scale invariant power
spectrum and Gaussian statistics. Even if non-linear effects in the evolution of perturbations are
expected, a simple calculation~\cite{fu02}, confirmed by more detailed analysis~\cite{maldacena},
shows that it is not possible to produce large non-Gaussianity within single field inflation as
long as the slow-roll conditions are preserved throughout the inflationary stage. Deviations from
Gaussianity can be larger in, e.g., multi-field inflation scenarios~\cite{fu02,bu} and are thus
expected to give details on the inflationary era.

As far as scalar modes are concerned, the deviation from Gaussianity has been parameterized by a
(scale-dependent) parameter, $f_{\rm NL}$. Various constraints have been set on this parameter,
mainly from CMB analysis~\cite{komatsu} (see Ref.~\cite{NGreview} for a review on both theoretical
and observational issues). Deviation from Gaussianity in the CMB can arise from primordial
non-Gaussianity, i.e. generated during inflation, post-inflation dynamics or radiation
transfer~\cite{transfert}. It is important to understand them all in order to track down the origin of
non-Gaussianity, if detected.

Among the other signatures of  non-linear dynamics is the fact that the Scalar-Vector and Tensor
(SVT) modes of the perturbations are no longer decoupled. This implies in particular that scalar
modes can generate gravity waves. Also, vector modes, that are usually washed out by the
evolution, can be generated. In particular, second-order scalar perturbations in the
post-inflation era will also contribute to B-mode polarization~\cite{SMoll} or
to multipole coupling in the CMB~\cite{prunet}, and it is thus
important to understand this coupling in detail.

In this article, we focus on the gravitational waves generated from scalar modes via second
order dynamics. Second-order perturbation theory has been investigated in various
works~\cite{Tomita,Mena,Matarr,2ndorder,clarkson1,langlois,ananda,nakamura,MAB} and a fully gauge-invariant approach to the problem was
recently given in Ref.~\cite{nakamura}. Second-order perturbations during inflation have also been
considered in Refs.~\cite{Acq,maldacena}, providing the prediction of the bispectrum of perturbations from inflation.

Two main  formalisms have been developed to study perturbations, and hence second order effects:
the $1+3$ covariant formalism~\cite{EB} in which exact gauge-invariant variables describing the physics of interest are first identified and exact equations describing their time and space evolution are then  derived and approximated with respect to the symmetry of the background to obtain results at the desired order, and  the coordinate based approach of Bardeen~\cite{Bardeen} in which gauge-invariants are identified by combining the metric and matter perturbations and then equations are found for them at the appropriate order of the calculation. In this article we carry out a detailed comparison of the two approaches up to second order, highlighting the advantages and disadvantages of each method, thus extending earlier work on the linear theory~\cite{BDE}. Our paper also extends the work of Ref.~\cite{langlois}, in which the relation between the two formalisms on super-Hubble scales is investigated. In particular, we show that the degree of success of one formalism over the other depends on the problem being addressed. This is the first time a complete and transparent matching of tensor perturbations in the two formalisms at first and second order is presented. We also show, using an analytical argument, that the power-spectrum of gravitational waves from second-order effects is much smaller than the first order on super-Hubble scales. This is in contrast to the fact that during the radiation era the generation of GW from primordial density fluctuations can be large enough to be detected in principle, though this requires the inflationary background of GW to be sufficiently
small~\cite{ananda}.

This paper is organized as follows. We begin by reviewing scalar field dynamics in
Section II within the 1 + 3 covariant approach. In Section III, we formulate the problem
within the covariant approach followed by a
reformulation in the coordinate approach in
Section IV. A detailed comparison of the two
formalisms is then presented in Section V. In Section VI, we study gravitational waves that are generated during
the slow-roll period of inflation. In particular, we introduce a generalization of the $f_{NL}$ parameter to take into account gravity waves and we compute the three point correlator
involving one graviton and two scalars. Among all three point functions involving scalar
and tensor modes, this correlator and the one involving three scalars are the dominant~\cite{maldacena}. Finally, we conclude in Section VII.

%%%%%%%%%%%%%%%%%%%%%%%%%%%%%%%%%%%
\section{Scalar field dynamics}\label{sec2}
%%%%%%%%%%%%%%%%%%%%%%%%%%%%%%%%%%%

Let us consider a minimally coupled scalar field with Lagrangian density\footnote{We use
conventions of Ref.~\cite{bi:wald}. Units in which $\hbar=c=k_{B}=1$ are used throughout this article, Latin indices $a,b,c...$ run from 0 to 3, whereas Latin
indices $i,j,k...$ run from $1$ to $3$. The symbol $\nabla$ represents the usual covariant
derivative and $\partial$ corresponds to partial differentiation. Finally the Hilbert-Einstein
action in presence of matter is defined by
\begin{equation}\label{action0}
{\cal A}=\int d x^{4} \sqrt{-g}\left[\sfrac{1}{16\pi G}R+ \ls_{\phi}\right]\;.
\end{equation}}
\be
\ls_{\phi}=-\sqrt{-g}\left[\case{1}/{2}\na_a\phi\na^a\phi +V(\phi)
\right]\;,
\ee
where $V(\phi)$ is a general (effective) potential expressing the self
interaction of the scalar field\iffalse; $\na_a$ is the covariant derivative
with respect to the metric $g_{ab}$\footnote{We shall assume in the
following that $\na_a$ acts on the  first argument on its right only
[e.g., $\na^e T^{a\cdots b}{}_{c\cdots d}\na_f Q^{g\cdots
h}{}_{l\cdots m} =(\na^e T^{a\cdots b}{}_{c\cdots d})(\na_f Q^{g\cdots
h}{}_{l\cdots m})$]. When $\na_a$ acts on a group of arguments, this
will be enclosed in parenthesis. }\fi.
The equation of motion for the field $\phi$  following from
$\ls_{\phi}$ is the Klein\hs Gordon equation
\be
\na_a\na^a\phi -V'(\phi)=0\;, \label{eq:KG}
\ee
where the prime indicates a derivative with respect to $\phi$.
The energy\hs momentum tensor of $\phi$ is of the form
\be
T_{ab}=\na_a\phi\na_b\phi -g_{ab}\left[\case{1}/{2}\na_c\phi\na^c\phi
+V(\phi)\right]\;; \label{eq:emt}
\ee
provided $\phi_{,a} \neq 0$, equation (\ref{eq:KG}) follows from the
conservation equation
\be
\na_bT^{ab}=0\;. \label{eq:cons}
\eei
We shall now assume that in the open region $U$ of spacetime that we
consider, the {\it momentum density} $\na^a\phi$ is {\it timelike}:
\be
\na_a\phi\na^a\phi<0\;. \label{eq:ass1}
\eei
This requirement implies two features: first, $\phi$ is not constant
in $U$, and so $\{\phi = const.\}$ specifies well-defined surfaces in
spacetime. When this is not true (i.e., $\phi$ is constant in $U$),
then by (\ref{eq:emt}),
\be
\na_a\phi = 0 ~~ \Leftrightarrow ~~T_{ab} = - g_{ab} V(\phi)\,
                    ~~\Rightarrow ~~V = \text{const.} \,, \label{eq:sf1a}
\ee
in $U$, [the last being necessarily true due to the conservation law (\ref{eq:cons}))] and we have an
effective cosmological constant in $U$ rather than a dynamical scalar field.

%.%%%%%%%%%%%%%%%%%%%%%%%%%%%%%%%%
\subsection{Kinematical quantities}\label{sec2A}
%%%%%%%%%%%%%%%%%%%%%%%%%%%%%%%%%

Our aim is to give a formal description of the scalar field in terms of fluid quantities;
therefore, we assign a 4-velocity vector $u^a$ to the scalar field itself. This will allow us to
define the dot derivative, i.e. the {\em proper time} derivative along the flow lines:
$\dot{T}^{a\cdots b}{}_{c\cdots d}\equiv u^e \na_eT^{a\cdots b}{}_{c\cdots d}$. Now, given the
assumption (\ref{eq:ass1}), we can choose the 4-velocity field $u^a$ as the unique timelike vector
with unit magnitude ($u^a u_a=-1$) parallel to the normals of the hypersurfaces $\{\phi=const.\}$
\, \cite{bi:madsen} \footnote{In the case of more than one scalar field, this choice can still be made 
for each scalar field 4-velocity, but not for the 4-velocity of the total fluid. A number of frame choices exist for the 4-velocity of the total fluid, the most common being the energy frame, where the total energy flux vanishes (see \cite{DBE} for a detailed description of this case.}
, \ber u^a&\equiv& -\inps \na^a\phi\;,
%~~~u^a u_a=-1\;,~~~\psi \equiv \dot{\phi}\nonumber\\
%&=& u^a\na_a\phi = (-\na_a\phi \na^a\phi)^{1/2}\;,
\label{eq:u}
\ear
where we have defined the field $\psi=\dot{\phi}= (-\na_a\phi \na^a\phi)^{1/2}$ to denote the
magnitude of the momentum density (simply momentum from now on). The choice
(\ref{eq:u}) defines $u^a$ as the unique timelike eigenvector of the
energy\hs momentum tensor (\ref{eq:emt}).\footnote{The quantity
$\psi$ will be positive or negative depending on the initial
conditions and the potential $V$; in general $\phi$ could oscillate
and change sign even in an expanding phase, and the determination of
$u^a$ by (\ref{eq:u}) will be ill\hs defined on those surfaces where
$\na_a\phi=0 \Rightarrow \psi=0$ (including the surfaces of maximum
expansion in an oscillating Universe). This will not cause us a
problem however, as we assume the solution is differentiable and
(\ref{eq:ass1}) holds almost everywhere, so determination of $u^a$
almost everywhere by this equation will extend (by continuity) to
determination of $u^a$ everywhere in $U$. }

The kinematical quantities associated with the ``flow vector'' $u^a$ can be obtained by a standard
method \cite{bi:ellis1,lang-vern}. We can define a projection tensor into the tangent 3-spaces
orthogonal to the flow vector: \be h_{ab}  \equiv g_{ab}+u_au_b\; \Rightarrow h^a{}_b
h^b{}_c=h^a{}_c\;, ~h_{ab}u^b=0\; ; \ee with this we decompose the tensor $\na_b u_a$ as \be \na_b
u_a=\n_b u_a-\acc_a u_b\;, ~~~\n_b u_a=\case{1}/{3}\Theta h_{ab} +\sigma_{ab}\;,  \label{eq:dec}
\ee where $\n_a$ is the spatially totally projected covariant derivative operator orthogonal to
$u^a$ (e.g., $\n_a f= h_a{}^b\na_b f$; see the Appendix of Ref.~\cite{bi:BDE} for details),
$\acc_a$ is the acceleration ($\acc_bu^b=0$), and $\sigma_{ab}$ the shear ($\sigma^a{}_a =
\sigma_{ab}u^b=0).$ Then the expansion, shear and acceleration are given in terms of the scalar
field by \ber \Theta& = & -\na_a(\inps\na^a\phi)
=-\inps\left[V'(\phi)+\dot{\psi}\right], \label{eq:exp} \\
\sigma_{ab} & = & - \inps\,h_{\lgl a}{}^c h_{b\rgl}{}^d \na_{c} \left[
\na_{d}\phi \right],
%+ \case{1}/{3} h_{ab} \na_c (\inps\na^c\phi)
\label{eq:shear}\\
a_a & = & -\inps\,\n_a\psi=-\inps(\na_a\psi +u_a\dot{\psi})\;,
\label{eq:acc}
 \ear
where the last equality in Eq.~(\ref{eq:exp}) follows on using the Klein\hs Gordon equation
(\ref{eq:KG}). We can see from Eq.~(\ref{eq:acc}) that $\psi$ is an {\em acceleration potential}
for the fluid flow \cite{bi:ellis2}. Note also that the vorticity vanishes:
 \be
 \omega_{ab}=-h_a{}^c h_b{}^d\na_{[d}\left(\inps\na_{c]}\phi\right)=0\;, \label{eq:vort}
 \ee an obvious result with the
choice (\ref{eq:u}), so that $\n_a$ is the covariant derivative operator in the 3-spaces
orthogonal to $u^a$, i.e. in the surfaces $\{\phi=\text{const.}\}$. As usual, it is useful to introduce a
scale factor $a$ (which has dimensions of length) along each flow\hs line by
 \be
 \frac{\dot{a}}{a}\equiv \case{1}/{3} \Theta = H\;,
 \ee
where $H$ is the usual Hubble parameter if the Universe is homogeneous and isotropic. Finally, it
is important to stress that
 \be
 \n_a\phi=0 \label{eq:gfi}
 \ee
which follows from our choice of $u^a$ via equation (\ref{eq:u}), a result that will be important for
the choice of gauge invariant (GI) variables and for the perturbations equations.

%%%%%%%%%%%%%%%%%%%%%%%%%%%%%%%%%%%%%%%%%%
\subsection{Fluid description of a scalar field}\label{sec2B}
%%%%%%%%%%%%%%%%%%%%%%%%%%%%%%%%%%%%%%%%%%

It follows from our choice of the four velocity (\ref{eq:u}) that we can represent a minimally
coupled scalar field as a perfect fluid; the energy\hs momentum tensor (\ref{eq:emt}) takes the
usual form for perfect fluids
\be T_{ab}=\mu u_a u_b +p h_{ab}\;, \label{eq:pf} \ee
where the energy
density $\mu$ and pressure $p$ of the scalar field ``fluid'' are given by
\ber
\mu &=& \case{1}/{2}\psi^2 +V(\phi)\;, \label{eq:ed}\\
p&=& \case{1}/{2}\psi^2 -V(\phi)\;. \label{eq:p} \ear
If the scalar field is not minimally
coupled this simple representation is no longer valid, but it is still possible to have an
imperfect fluid form for the energy\hs momentum tensor\,\cite{bi:madsen}.

Using the perfect fluid energy\hs momentum tensor (\ref{eq:pf}) in
(\ref{eq:cons}) one obtains the energy and momentum conservation
equations
\ber
 \dot{\mu} +\psi^2\Theta&=&0\;,\label{eq:encon} \\
\psi^2\acc_a+\n_ap&=&0 \;. \label{eq:momcon} \ear If we now substitute $\mu$ and $p$ from
Eqs.~(\ref{eq:ed}) and (\ref{eq:p}) into Eq.~(\ref{eq:encon}) we obtain the 1+3 form of the
Klein\hs Gordon equation (\ref{eq:KG}): \be \ddot{\phi} +\Theta \dot{\phi} +V'(\phi)=0\;,
\label{eq:KG2} \ee an exact ordinary differential equation for $\phi$ in any space\hs time with
the choice (\ref{eq:u}) for the four\hs velocity. With the same substitution,
Eq.~(\ref{eq:momcon}) becomes an identity for the acceleration potential $\psi$. It is convenient
to relate $p$ and $\mu$ by the {\em index $\gamma$} defined by \be
 \label{eq:e6}
p = ( \gamma - 1) \mu  ~~\Leftrightarrow ~~ \gamma = \frac{p + \mu}{\mu} ~=~\frac{\psi^2}{\mu}\;.
\ee This index would be constant in the case of a simple one-component fluid, but in general will
vary with time in the case of a scalar field: \be
\frac{\dot{\gamma}}{\gamma}=\Theta(\gamma-2)-2\frac{V'}{\psi}\;. \label{eq:dotgamma} \ee Finally,
it is standard to {\it define} a speed of sound as\,
 \be
 c_s^2\equiv\frac{\dot{p}}{\dot{\mu}} =
 \gamma-1 -\frac{\dot{\gamma}}{\Theta\gamma}\;. \label{eq:sos}
 \ee

%%%%%%%%%%%%%%%%%%%%%%%%%%%%%%%%%%%%%%%%
\subsection{Background equations}\label{sec2C}
%%%%%%%%%%%%%%%%%%%%%%%%%%%%%%%%%%%%%%%%

The previous equations assume nothing on the symmetry of the spacetime. We now specify it further
and assume that it is close to a flat Friedmann-Lema\^{\i}tre spacetime (FL), which we consider as our
background spacetime. The homogeneity and isotropy assumptions imply that

\be
\sigma^2 \equiv \frac{1}{2}\sigma_{ab}\sigma^{ab}=0\;,\,\,\omega_{ab}= 0\;, ~~\n_a f=0\;,
\label{eq:rwcond1}
\ee
where $f$ is any scalar quantity; in particular
\be
\n_a\mu =\n_a p=0 ~~\Rightarrow~~\n_a \psi = 0\,, ~a_a = 0\;.
\ee
The background (zero\hs order) equations are given by \cite{bi:scalar}:
\ber
 && 3\dot{H} + 3H^2   =  8\pi G\left[V(\phi) - \psi^2\right] \, ,\label{eq:e1b}  \\
 &&3H^2     =    8\pi G\left[\sfrac12 \psi^2 + V(\phi)\right]  \, \label{eq:e1a}  \\
  &&\dot{\psi} + 3H\psi + V'(\phi)= 0\; \Leftrightarrow
  \dot{\mu} + 3H\psi^2 = 0     \; ,\label{eq:e1c}
\ear
where all variables are a function of cosmic time $t$ only.

\section{gravitational waves from density perturbations: covariant formalism}

%%%%%%%%%%%%%%%%%%%%%%%%%%%%%%%%%%%%%%%%
\subsection{First order equations}\label{sec2D}
%%%%%%%%%%%%%%%%%%%%%%%%%%%%%%%%%%%%%%%%

The study of linear perturbations of a FL background are relatively straightforward. Let us
begin by defining the \emph{first-order gauge-invariant} (FOGI) variables
corresponding respectively to the spatial fluctuations in the energy density,
expansion rate and spatial curvature:
 \ber
 X_a&=&\n_a\mu\;,\nonumber \\
 Z_a&=&\n_a\Theta\;,\nonumber\\
 C_a&=&a^3\n_a\tilde{R}\;.
 \ear
The quantities are FOGI because they vanish exactly in the background FL spacetime
\cite{bi:SW,bi:BS}.  It turns out that a more suitable quantity for describing density
fluctuations is the co-moving gradient of the energy density:
 \be
 \da= \frac{a}{\mu}X_a\;,
 \ee
where the ratio $X_a/\mu$ allows one to evaluate the magnitude of the energy density perturbation
relative to its background value and the scale factor $a$ guarantees that it is dimensionless and
co-moving.

These quantities exactly characterize the inhomogeneity of any fluid;
however we specifically want to characterize the inhomogeneity of the
scalar field: this cannot be done using the spatial gradient
$\n_a\phi$ because it identically vanishes in any space-time by virtue of our
choice of 4-velocity field $u^a$.  It follows that in our
approach the inhomogeneities in the matter field are completely
incorporated in the spatial variation of the momentum density: $\n_a\psi$, so
it makes sense to define the dimensionless gradient
 \be
 \Psi_a\equiv \frac{a}{\psi}\n_a\psi\;,
 \label{eq:psidef}
 \ee
which is related to $\da$ by
 \be
 \da= \frac{\psi^2}{\mu}\Psi_a=\gamma\Psi_a\;,
 \ee
where we have used Eq.~(\ref{eq:ed}) and $\gamma$ is given by Eq.~(\ref{eq:e6}); comparing
Eq.~(\ref{eq:psidef}) and  Eq.~(\ref{eq:acc}) we see that $\Psi_a$ is proportional to the
acceleration: it is a gauge-invariant
 measure of  the spatial variation of proper time along the flow lines of $u^a$
 between two surfaces $\phi=$const. (see Ref.~\cite{bi:ellis1}).
The set of linearized equations satisfied by the FOGI variables consists of the
\emph{evolution equations}
 \ber
&&\dot X_a=-4H X_a-\psi^2 Z_a,\\
&&\dot Z_a=-3H Z_a- 4 \pi G X_a+\n_a\div \acc,\\
&&\dot \sigma_{ab}-\n_{\lgl a}\acc_{b\rgl} =-2H\sigma_{ab}-E_{ab}, \\
&&\dot E_{ab}-\curl H_{ab}=-4 \pi G\psi^2\sigma_{ab}-\Theta E_{ab},\\
&&\dot H_{ab}+\curl E_{ab}=-3H H_{ab} \label{eq:dotH};
\ear
and the \emph{constraints}

\ber
0&=&\sfrac{8 \pi G}{3} X_a-\div E_a,\\
0&=&\sfrac{2}{3} Z_a-\div \sigma_a,\label{e41}\\
0&=&\div H_a,\\
0&=&H_{ab}-\curl\sigma_{ab}\;,\label{eq:H}\\
0&=&\curl X_a,\\
0&=&\curl Z_a\;. \ear The $\curl$ operator is defined by
$\curl\psi_{ab}=(\curl\psi)_{ab}=\varepsilon_{cd\langle a}\n^c\psi_{b\rangle}^d$ where
$\epsilon_{abc}$ is the completely antisymmetric tensor with respect to the spatial section defined by $\epsilon_{bcd}=\epsilon_{abcd}u^{a}$, $\epsilon_{abcd}$ being the volume antisymmetric tensor such that $\epsilon_{0123}=\sqrt{-g}$. The divergence $\div$ of a rank $n$ tensor is a rank $n-1$ tensor
defined by $(\div \psi)_{i_1...i_{n-1}}\equiv \n^{i_n} \psi_{i_1...i_n}$.

Because the background is homogeneous and isotropic, each FOGI
vector may be uniquely split into a \emph{curl-free} and
\emph{divergence-free} part, usually referred to as scalar and
vector parts respectively, which we write as
\be
V_a={V\S}_a+{V\V}_a\;,
\ee
where $\curl{V\S}_a=0$ and $\div{V\V}=0$.  Similarly, any tensor
may be invariantly split into scalar, vector and tensor parts:
\be
T_{ab}={T\S}_{ab}+{T\V}_{ab}+{T\T}_{ab}
\ee
where $\curl{T\S}_{ab}=0$ , $\div\div{T\V}=0$ and $\div{T\T}_a=0$.
It follows therefore that in the above  equations we can  separately equate scalar,
vector and tensor parts and obtain equations that independently characterize the evolution
of each type of perturbation. In the case of a scalar field, the vorticity is exactly zero, so there
is no vector contribution to the perturbations.

Let us now concentrate on scalar perturbations at linear order. It is clear from the above
discussion that pure scalar modes are characterized by the vanishing of the magnetic
part of the Weyl tensor:  $H_{ab}=0$, so the above set of equations reduce to a set of two
coupled differential equations for $X_a$ and $Z_a$:
\ber
 && \dot X_a + 4H X_a = -\psi^2 Z_a \label{eq:II-40}\\
 && \dot Z_a +  3H Z_a = -4 \pi G X_a-{\psi^{-2}}\n^2 X_a,\label{eq:II-41}
\ear and a set of coupled evolution and constraint equations that
determine the other variables \ber
  && \dot{\sigma}_{ab}  = - \psi^{-2}\n_{\lgl a} X_{b\rgl}- 2H \sigma_{ab}  -
      E_{ab}.\label{eq:II-42}\\
  && \dot{E}_{ab}  =  - 4 \pi G\,\psi^2\,\sigma_{ab}
 - 3H E_{ab},\label{eq:II-43} \\
  && \div\sigma_a = \sfrac{2}{3}\,Z_{a}, \label{eq:II-44}\\
  && \curl\sigma_{ab} = 0, \label{eq:II-45}\\
  &&  \div E_{a} = \sfrac{8 \pi G}{3}\,X_{a}\label{eq:II-46}.
\ear
%%%%%%%%%%%%%%%%%%%%%%%%%%%%%%%%%%%%%%%%
\subsection{Gravitational waves from density perturbations}\label{sec3E}
%%%%%%%%%%%%%%%%%%%%%%%%%%%%%%%%%%%%%%%%
The preceding discussion deals with first-order variables and their behavior at linear order. It
is important to keep in mind that we were able to set $H_{ab}=0$ only because
pure scalar perturbations in the absence of vorticity implies that
$\curl\sigma_{ab}=0$ at first order. The vanishing
of the magnetic part then follows from equation (\ref{eq:H}). However, at
second order $\curl\sigma_{ab}\neq0$. We denote the non-vanishing contribution at second order by~\cite{clarkson1}
\[\Sigma_{ab}=\curl\sigma_{ab}.\] The new variable is \emph{second-order and gauge-invariant} (SOGI), as it vanishes at all lower orders ~\cite{bi:SW}. It should be
noted that the new variable is just the magnetic part of the Weyl tensor subject to the conditions
mentioned above i.e. \be \Sigma_{ab}=H_{ab}|_{\omega=0}\label{eq:Sig}. \ee 
We are interested
in the properties inherited by the new variable from the magnetic part of the Weyl tensor. In
particular, it can be shown that the new variable is transverse and traceless at this order and is
thus a description of gravitational waves.
It should be stressed that in full generality, there are tensorial modes even at first order. By assuming that there are none, we explore a particular subset in the space of solutions. From the "iterative resolution" point of view, this means that we constrain the equations in order to focus on second order GWs sourced by terms quadratic in scalar perturbations. In doing so, we artificially switch off GW perturbations at first order. 

%%%%%%%%%%%%%%%%%%%%%%%%%%%%%%%%%%%%%%%%%%
\subsection{Propagation equation}
%%%%%%%%%%%%%%%%%%%%%%%%%%%%%%%%%%%%%%%%%%
The propagation of the new second-order
variable now needs to be investigated using a covariant set of
equations that are linearized to second-order about FL. We make
use of Eqs. (\ref{eq:encon}), (\ref{eq:momcon}) and the following
evolution equations which are up to second order in magnitude;
\begin{eqnarray}
\dot{E}_{ab} &=&-\Theta E_{ab}+\curl\Sigma_{ab}-4 \pi G\psi^{2}\sigma_{ab}+3\sigma_{c\langle a}{E_{b\rangle}}^{c}\label{eq:dotESogi},\\
\dot{\Sigma}_{ab} &=&-\Theta \Sigma_{ab}-\curl
E_{ab}-2\epsilon_{cd\langle
a}\acc^{c}{E_{b\rangle}}^{d}\label{eq:SigmaSOGI},
\end{eqnarray}
together with the constraint
\begin{equation}\label{eq:mcons}
 \acc^{a}=-{\psi^{-2}}\tilde{\na}^{a}p=-\frac{3}{8 \pi G \psi^{2}}\div E^{a}.
\end{equation}
Unlike at first-order, where the splitting of tensors into their scalar, vector and tensor parts
is possible, at second order this can only be achieved for SOGI variables.

We may isolate the tensorial part of the equations by decoupling $\Sigma_{ab}$: since it is
divergence free it is already a pure tensor mode, whereas $E_{ab}$ is not. The wave equation for the
gravitational wave contribution can be found by first taking the time derivative of
(\ref{eq:SigmaSOGI}) and making appropriate substitutions using the evolution equations and
keeping terms up to second order. The wave equation for $\Sigma_{ab}$ then reads:
 \be
 \ddot{\Sigma}_{ab}-\n^{2}\Sigma_{ab}+7H\dot{\Sigma}_{ab}+(12H^2-16 \pi G\psi^{2})\Sigma_{ab}=S_{ab}\label{eq:dotdotH}\ee
 where the source is given by the cross-product of the electric-Weyl curvature and its divergence (or acceleration):
\iffalse \begin{eqnarray} S_{ab}&=&-4\left(\frac{1}{\psi^{2}}\epsilon_{cd\langle
 a}{{E_{b\rangle}}^{d}}\div(E)^{c}\right)^{.}\nonumber\\&-&\left(32H
 -30H c^{2}_{s}\right)\left(\frac{1}{\psi^{2}}\epsilon_{cd\langle
  a}{E_{b\rangle}}^{d}\div(E)^{c}\right),
\label{eq:Source}
\end{eqnarray}\fi
\ber
S_{ab}&=&-\left[2u^{e}\nabla_{e}+16H-15Hc_s^{2}
 \right]\nonumber\\&&~~~~
 \left(\frac{1}{4 \pi G\psi^{{2}}}\epsilon_{cd\langle
 a}{E_{b\rangle}}^{d}\div E^{c}\right).
 \label{eq:Source}
\ear
To obtain this, we have used the fact that with a flat background space-time
\ber
\curl\curl
T_{ab}&=&-\n^{2}T_{ab}+\frac{3}{2}\n_{\langle a}\div T_{ b\rangle}
\label{eq:III-5}\nonumber
\ear
and used the commutation relation
\begin{eqnarray*}
 (\curl T_{ab})^{.} &=&
\curl\dot{T}_{ab}-\frac{1}{3}\Theta \curl T_{ab}-\epsilon_{cd\langle
a}\sigma^{ec}D_{|e|}{T_{b\rangle}}^{d}\\&&+\epsilon_{cd\langle
a}[\acc^{c}{\dot{T}_{b}}^{d}+\frac{1}{3}\Theta\acc^{c}{{T}_{b\rangle}}^{d}].
\end{eqnarray*}
 We have also used Eqs.
(\ref{eq:dotgamma}) and (\ref{eq:sos}) to eliminate $\dot{\psi}/\psi$ from the source term. It can
also be shown that $S_{ab}$ is transverse, illustrating that Eq.~(\ref{eq:dotdotH}) represents the gravitational wave contribution at second order. Note that this is a local description of gravitational waves, in contrast to the non-local extraction of tensor modes by
projection in Fourier space. Since $\Sigma_{ab}$ contains exactly the correct number of degrees of freedom possible in GW, any other variable we may choose to describe GW must be related by quadrature, making this a suitable master variable. The situation is analogous to the description of electromagnetic waves: Should we use the vector potential, the electric field, or the magnetic field for their description? Mathematically it doesn't matter of course -- each variable obeys a wave equation and the others are related by quadrature. Physically, however, it's the electric and magnetic fields which drive charged particles through the Lorentz force equation~-- the electromagnetic analogue of the geodesic deviation equation. 

In order to express the gravitational wave equation in Fourier space, we define our normalised
tensor harmonics as
\begin{equation}
 Q^{ab}=\frac{\xi^{ab}}{(2\pi)^{3/2}}e^{i\bk\cdot\bx} ,
\end{equation}
where $\xi^{ab}$ is the polarization tensor, which satisfies the (background) tensor Helmholtz equation:
$\tilde{\na}^{2}Q_{ab}=-(\emph{q}^{2}/a^{2})Q_{ab}$. As $q_a$ is required to satisfy $q_a u^a=0$ in the background, it can thus be identified with a 3-vector and will subsequently written in bold when necessary. We denote harmonics of the opposite
polarization with an overbar. Amplitudes of $\Sigma_{ab}$ may be extracted via
\begin{equation}
 \Sigma(\bk,t)= \int{\dd^{3}\bk\left[\Sigma_{ab}(\bx,t)Q^{*ab}(\bk,\bx)\right]},
\end{equation}
with an analogous formula for the opposite parity. This implies that our original variable may be
reconstructed from
\begin{equation}
 \Sigma_{ab}=\int\dd^{3}\bk\left[{\Sigma}(\bk,t)Q_{ab}(\bk,\bx)
+\bar{\Sigma}(\bk,t)\bar Q_{ab}(\bk,\bx)\right].
\end{equation}
The same relations hold for any transverse tensor. Hence, our wave equation in Fourier space is
\ber &&\Sigma^{\prime\prime}(\bk,\eta)+6\mathcal{H}\Sigma^{\prime}(\bk,\eta)+
[k^2+12\mathcal{H}^2-16 \pi G\psi^{2}]\Sigma(\bk,\eta)
  \nonumber\\ &&~~~~~~~~~~~~~~~~~~~~~~~~~~~~~~~~~~~~ = S(\bk,\eta),\label{eq:wave}
\ear
with an identical equation for the opposite polarization. We have converted to conformal time $\eta$, where a prime denotes derivatives with respect to $\eta$, and we have defined the conformal Hubble parameter as $\mathcal{H}=a'/a$. The source term is composed
of a cross-product of the electric part of the Weyl tensor and its
divergence. At first-order, the electric Weyl tensor is a pure scalar mode, and can therefore be expanded in terms of scalar harmonics.
To define these, let $Q^{(s)}=e^{i\bq\cdot\bx}/(2\pi)^{3/2}$, be a
solution to the Helmholtz equation:
$\tilde{\na}^{2}Q^{(s)}=-(\emph{q}^{2}/a^{2})Q^{(s)}.$ Beginning
with this basis, it is possible to derive vectorial and (PSTF) tensorial
harmonics by taking successive spatial derivatives as follows:
 \begin{eqnarray}
  Q_{a}^{(s)}&=&\tilde{\na}_a Q^{(s)}= i\frac{
\emph{q}_{a}}{a} Q^{(s)}, \\
Q_{ab}^{(s)}&=&\tilde{\na}_{\langle
a}\tilde{\na}_{b\rangle}Q^{(s)}=-a^{-2}\left({\emph{q}_{a}\emph{q}_{b}}
-\frac{1}{3}h_{ab}{\emph{q}^{2}}\right)Q^{(s)}.\nonumber\\
\end{eqnarray}
This symmetric tensor has the additional property
$q^{a}q^{b}Q_{ab}^{(s)}=-(2\emph{q}^{4}/{3a^{2}})Q^{(s)}.$ Using this representation we can
express our source in Eq.~(\ref{eq:wave}) in terms of a convolution in Fourier space, by expanding
the electric Weyl tensor as 
\be E(\bq,\eta)=\int\dd^{3}\bx
E_{ab}Q_{(S)}^{*ab}(\bq,\bx). 
\ee 
Then, the right hand side of Eq.~(\ref{eq:Source}) expressed in conformal time, accompanied by
appropriate Fourier decomposition of each term and making use of the normalization condition for
orthonormal basis, yields:
\begin{widetext}
\begin{eqnarray}S(\bk,\eta)&=&\int\dd^3\bq
A(\bq,\bk)\left\{2\left[E(\bq,\eta)E(\bk-\bq,\eta)\right]'+(16-15c^{2}_{s})
\mathcal{H}\,E(\bq,\eta)E(\bk-\bq,\eta)\right\}
\end{eqnarray}
where
\be
A(\bq,\bk)=\frac{i}{6 \pi G a^{3}\psi^{2}}\epsilon_{cd\langle
a}q_{b\rangle}q^{d}(k^{c}-q^{c})|\bk-\bq|^{2}\xi^{ab}(\bk),
\ee
with a similar expression for the other polarization.
%\end{widetext}

In principle we can now solve for the gravitational wave contribution $\Sigma_{ab}$, and calculate the power spectrum of gravitational waves today. For this however, we need initial conditions for the electric Weyl tensor (or, alternatively $\Psi_{a}$).

\section{gravitational waves from density perturbations: coordinate based approach}\label{sec4}
%%%%%%%%%%%%%%%%%%%%%%%%%%%%%%%%%%%

In this formalism, we consider perturbations around a FL universe
with Euclidean spatial sections and expand the metric as
%\begin{widetext}
\begin{equation}\label{metric}
 \dd s^2=
 a^2(\eta)\left\{-(1+2A)\dd\eta^2 + 2\omega_i\dd x^i\dd\eta+
 \left[(1+2 C)\delta_{ij}+ h_{ij}\right]\dd x^i\dd x^j\right\}\,,
\end{equation}
\end{widetext}
where $\eta$ is the conformal time and $a$ the scale factor. We perform a scalar-vector-tensor
decomposition as
\begin{equation}\label{eq:I-2}
 \omega_{i}=D_i B + \bar B_i\,,
\end{equation}
and
\begin{equation}\label{eq:I-3}
 h_{ij}=2 \bar {\mathcal E}_{ij} + D_i \bar {\mathcal E}_j + D_j \bar {\mathcal E}_i + 2 D_iD_j {\mathcal E}\;,
\end{equation}
where $\bar B_i$, $\bar {\mathcal E}_i$ are transverse ($D_i \bar {\mathcal E}^i= D_i \bar B^i$),
and $\bar {\mathcal E}_{ij}$ is traceless and transverse ($\bar {\mathcal E}^i_i=D_i \bar
{\mathcal E}^i_j=0$). Latin indices $i,j,k...$ are lowered by use of the spatial metric, e.g.
$B^i=\gamma^{ij}B_j$. We fix the gauge and work in the Newtonian gauge defined by $B_i={\mathcal
E}=B=0$ so that $\Phi=A$ and $\Psi=-C$ are the two Bardeen potentials.
As in the previous sections, we assume that the matter content is a scalar field $\phi$ that can
be split into background  and perturbation contributions:  $\phi=\phi(\eta)+\delta\phi(\eta,\bx)$. The gauge invariant scalar
field perturbation can be defined by
\begin{equation}
 Q \equiv \delta\phi -\phi'\frac{C}{\cal{H}}\;,
\end{equation}
where ${\cal{H}}\equiv a'/a \equiv a H$. We denote the field perturbation in Newtonian gauge by $\chi$ so
that $Q=\chi+(\phi'/\mathcal{H})\Psi$. Introducing
\begin{equation}
\varepsilon=\frac{3}{2}\frac{\phi'^2}{\mu}\;,
\end{equation}
the equation of state~(\ref{eq:e6}) takes the form $\gamma=w+1=2\varepsilon/3$. We thus have two expansions: one concerning the perturbation of the metric and the other in the slow-roll parameter $\epsilon$.

%%%%%%%%%%%%%%%%%%%%%%%%%%%%%%%%%%%%%
\subsection{Scalar modes}
%%%%%%%%%%%%%%%%%%%%%%%%%%%%%%%%%%%%%

Focusing on scalar modes at first order in the perturbation, it is convenient to introduce
\begin{equation}
 v=aQ
\end{equation}
and
\begin{equation}
 z\equiv a\frac{\phi'}{\cal{H}},
\end{equation}
in terms of which the action~(\ref{action0}) takes the form
\begin{equation}\label{actionscal1}
 S_{\rm scal}=\frac{1}{2}\int\dd^3\bx\dd\eta\left[
 (v')^2-(\partial_iv)^2+\frac{z''}{z}v^2
 \right]\;,
\end{equation}
when expanded to second order in the perturbations. It is the action of a canonical scalar field with effective square mass $m_v^2=-z''/z$. $v$ is the canonical variable that must be quantized~\cite{bmf}. It is decomposed as follows
\begin{equation}
 \hat v(\bx,\eta) = \int\frac{\dd^3\bk}{(2\pi)^{3/2}}
 \left[v_k(\eta)\hbox{e}^{i\bk.\bx}\hat{a}_{\bk} +
   v_k^*(\eta)\hbox{e}^{-i\bk.\bx}\hat a_{\bk}^\dag
 \right]\;.
\end{equation}
Here $v_k$ is solution of the Klein-Gordon equation
\begin{equation}\label{kgv}
v_k''+\left(k^2-\frac{z''}{z}\right)v_k=0
\end{equation}
and the annihilation and creation operators satisfy the commutation relation, $[\hat a_{\bk}, \hat
a_{\bk'}^\dag]=\delta(\bk-\bk')$. We define the free vacuum state by the requirement $\hat
a_{\bk}|0\rangle=0$ for all $\bk$.

From the Einstein equation, one can get the expression for the Bardeen potential (recalling that
$\Psi=\Phi$)
\begin{equation}
 \Delta\Phi = 4\pi G\frac{\phi^{\prime2}}{\cal{H}}\left(\frac{v}{z}\right)',
 \quad
 \left(\frac{a^2\Phi}{\mathcal{H}}\right)'=4\pi G zv
\end{equation}
and for the curvature perturbation in comoving gauge
\begin{equation}
 \mathcal{R}=-v/z.
\end{equation}
Once the initial conditions are set, solving Eq.~(\ref{kgv}) will give the evolution of
$v_k(\eta)$ during inflation, from which $\Phi_k(\eta)$ and ${\cal R}_k(\eta)$ can be deduced,
using the previous expressions.

Defining the power spectrum as
\begin{equation}
 \langle{\cal R}_{\bk}{\cal R}^*_{\bk'}\rangle = \frac{2\pi^2}{k^3}{\cal P}_{\cal
 R}(k)\delta^{(3)}(\bk-\bk'),
\end{equation}
one easily finds that
\begin{equation}\label{PR}
 {\cal P}_{\cal R}(k) = \frac{k^3}{2\pi^2}\left|\frac{v_k}{z}\right|^2.
\end{equation}
Note also that $z$ and $\varepsilon$ are related by the simple relation
\begin{equation}\label{za}
 \sqrt{4\pi G}\,z=a\sqrt{\varepsilon}\;,
\end{equation}
so that
\begin{equation}\label{chiQ}
 \chi=Q-\frac{z}{a}\Phi=Q-\sqrt{\frac{\varepsilon}{4\pi G}}\Phi.
\end{equation}

%%%%%%%%%%%%%%%%%%%%%%%%%%%%%%%%%%%%
\subsection{Gravitational waves at linear order}
%%%%%%%%%%%%%%%%%%%%%%%%%%%%%%%%%%%%

At first order, the tensor modes are gauge invariant and their propagation equation is given by
\begin{equation}\label{Ebar1}
 \bar {\mathcal E}_{ij}'' + 2{\cal{H}}\bar {\mathcal E}_{ij}'-\Delta\bar {\mathcal E}_{ij}=0
\end{equation}
since a minimally coupled scalar field has no anisotropic stress. Defining the reduced variable
\begin{equation}\label{mu2E}
 \mu_{ij}= \frac{a}{\sqrt{8\pi G}}\bar {\mathcal E}_{ij}\;,
\end{equation}
the action~(\ref{action0}) takes the form
\begin{equation}\label{actiontens1}
 S_{\rm tens}=\frac{1}{2}\int\dd^3\bx\dd\eta\left[
 (\mu'_{ij})^2-(\partial_k\mu_{ij})^2+\frac{a''}{a}(\mu_{ij})^2
 \right]
\end{equation}
when expanded tosecond order. Developing $\bar {\mathcal E}_{ij}$, and similarly $\mu_{ij}$, in
Fourier space:
\begin{equation}\label{decFou}
 \bar {\mathcal E}_{ij}=\sum_{\lambda=+,\times}\int\frac{\dd^3\bk}{(2\pi)^{3/2}}
 {\mathcal E}_\lambda \varepsilon_{ij}^\lambda(\bk) \hbox{e}^{i\bk.\bx},
\end{equation}
where $\varepsilon_{ij}^\lambda$ is the polarization tensor, the action~(\ref{actiontens1}) takes
the form of the action for two canonical scalar fields with effective square mass $m_\mu^2=-a''/a$
\begin{equation}\label{actiontens2}
 S_{\rm tens}=\frac{1}{2}\sum_\lambda\int\dd^3\bx\dd\eta\left[
 (\mu'_\lambda)^2-(\partial_k\mu_\lambda)^2+\frac{a''}{a}\mu_\lambda^2
 \right].
\end{equation}
If one considers the basis $({\bf e}_1,{\bf e}_2)$ of the 2 dimensional space orthogonal to $\bk$
then $\varepsilon^{\lambda}_{ij}=(e^1_ie^1_j-e^2_ie^2_j) \delta_+^\lambda + (e^1_ie^2_j +
e^2_ie^1_j)\delta_\times^\lambda$.

$\mu_\lambda$ are the two degrees of freedom that must be quantized~\cite{bmf} and we expand them
as
\begin{eqnarray}
 \hat \mu_{ij}(\bx,\eta) &=& \sum_\lambda\int\frac{\dd^3\bk}{(2\pi)^{3/2}}
 \left[\mu_{k,\lambda}(\eta)\hbox{e}^{i\bk.\bx}\hat{b}_{\bk,\lambda} \right.\nonumber\\
 &&\qquad\qquad\left.
   +\mu_{k,\lambda}^*(\eta)\hbox{e}^{-i\bk.\bx}\hat b_{\bk,\lambda}^\dag
 \right]\varepsilon_{ij}^\lambda(\bk) .
\end{eqnarray}
$\mu_k$ is solution of the Klein-Gordon equation
\begin{equation}\label{mueq}
 \mu_k''+\left(k^2 -\frac{a''}{a}\right)\mu_k=0,
\end{equation}
where we have dropped the polarization subscript. The annihilation and creation operators satisfy
the commutation relations, $[\hat b_{\bk,\lambda}, \hat b_{\bk',\lambda'}^\dag] = \delta(\bk-\bk')
\delta_{\lambda\lambda'}$ and $[\hat a_{\bk}, \hat b_{\bk',\lambda}^\dag]=0$. We define the free
vacuum state by the requirement $\hat b_{\bk,\lambda}|0\rangle=0$ for all $\bk$ and $\lambda$.

Defining the power spectrum as
\begin{equation}\label{specE1}
 \langle {\mathcal E}_{\bk,\lambda} {\mathcal E}^*_{\bk',\lambda'}\rangle = \frac{2\pi^2}{k^3}{\cal P}_T(k)
 \delta^{(3)}(\bk-\bk')\delta_{\lambda\lambda'},
\end{equation}
one easily finds that
\begin{equation}
 {\cal P}_T(k) = 16\pi G\frac{k^3}{2\pi^2}\left|\frac{\mu_k}{a}\right|^2,
\end{equation}
where the two polarizations have the same contribution.

%%%%%%%%%%%%%%%%%%%%%%%%%%%%%%%%%%%%%%%
\subsection{Gravitational waves from density perturbations}
%%%%%%%%%%%%%%%%%%%%%%%%%%%%%%%%%%%%%%%

At second order, we split the tensor perturbation as $\bar{\mathcal E}_{ij}= \bar{\mathcal
E}^{(1)}_{ij} + \bar{\mathcal E}^{(2)}_{ij}/2$. The evolution equations of $\bar {\mathcal
E}^{(2)}_{ij}$ is similar to Eq.~(\ref{Ebar1}),  but inherits a source term quadratic in the first
order perturbation variables and from the transverse tracefree (TT) part of the stress-energy tensor
\begin{equation}\label{Ebar2bis}
 a^2\left[T^i_j\right]^{\rm TT}= \gamma^{ip}\left[\partial_j\chi\partial_p\chi\right]^{\rm TT}.
\end{equation}
It follows that the propagation equation is
\begin{equation}\label{Ebar2}
 \bar{\mathcal E}^{(2)\prime\prime}_{ij} + 2{\cal{H}}{\bar{\mathcal E}^{(2)\prime}_{ij}}
 -\Delta\bar{\mathcal E}^{(2)}_{ij}=S^{\rm TT}_{ij}\;,
\end{equation}
where $S^{\rm TT}_{ij}$ is a TT tensor that is quadratic in the first
order perturbation variables.

Working in Fourier space, the TT part of any tensor can easily be extracted by means of the
projection operator
\begin{equation}\label{eq:III-2}
 \perp_{ij}(\unitbk) = \delta_{ij} - \unitk_i\unitk_j,
\end{equation}
where $\hat k^i=k^i/k$ (note that $\perp_{ij}(\unitbk)$ is not analytic in $k$ and is a non-local
operator) from which we get
\begin{eqnarray}\label{tensextra}
 S^{\rm TT}_{ij}(\bk,\eta) &=&
 \left[\perp_{i}^a\perp_{j}^b-\frac{1}{2}
 \perp_{ij}\perp^{ab}\right]S_{ab}(\bk,\eta)\nonumber\\
 &\equiv&P_{ij}^{ab}(\bk)S_{ab}(\bk,\eta).
\end{eqnarray}
The source term is now obtained as the TT-projection of the second order Einstein tensor quadratic
in the first order variables and of the stress-energy tensor
\begin{equation}\label{Ebar2b}
 S_{ab} = S^{(2)}_{ab,{\rm SS}} + S^{(2)}_{ab,{\rm ST}}
                 + S^{(2)}_{ab,{\rm TT}}.
\end{equation}
The three terms respectively indicate terms involving products of first order scalar quantities, first order
scalar and tensor quantities and first order tensor quantities. The explicit
form of the first term is
\begin{eqnarray}\label{sourceterm}
 S^{\rm TT}_{ij}&=&4\left[\partial_i\Phi\partial_j\Phi+ 4\pi G\,
 \partial_i\chi\partial_j\chi
  \right]^{\rm TT}.
\end{eqnarray}
The first term was considered in Ref.~\cite{gwpot} and the second term was shown to be the
dominant contribution for the production of gravitational waves during preheating~\cite{gwphi}. In
Fourier space, it is given by
\begin{eqnarray}\label{sourceSS}
 S^{(2)}_{ab,{\rm SS}} &=& -4\left[ \int \dd^3\bq \,
     q_bq_a\Phi(\bq,\eta)\Phi(\bk-\bq,\eta)\right.\nonumber\\
 && +\left. 4\pi G\int \dd^3\bq \,
     q_bq_a\chi(\bq,\eta)\chi(\bk-\bq,\eta)\right].
\end{eqnarray}
$\mu^{(2)}_{ij}(\bx,\eta)$ can be decomposed as in Eq.~(\ref{decFou}), using the same
definition~(\ref{mu2E}) at any order. The two polarizations evolve according to
\begin{equation}\label{mu2propa}
 \mu_\lambda^{(2)\prime\prime} + \left(k^2-\frac{a''}{a}\right) \mu_\lambda^{(2)} =
  -\frac{2a}{\sqrt{8\pi G}}  P_{ij}^{ab}S_{ab,{\rm SS}}^{(2)}\varepsilon_{\lambda}^{ij}.
\end{equation}
Since the polarization tensor is a TT tensor, it is obvious that $P_{ij}^{ab}
\varepsilon_{\lambda}^{ij} = \varepsilon_{\lambda}^{ab}$,  so that
\begin{widetext}
\begin{equation}\label{eqmu2}
 \mu_\lambda^{(2)\prime\prime} + \left(k^2-\frac{a''}{a}\right) \mu_\lambda^{(2)} =
  -\frac{4a}{\sqrt{8\pi G}} \varepsilon^{ij}_\lambda \int \dd^3\bq \,
  q_iq_j\left[\Phi(\bq,\eta)\Phi(\bq-\bk,\eta)+4\pi G\,\chi(\bq,\eta)\chi(\bq-\bk,\eta)
  \right].
\end{equation}
\end{widetext}

From the equation~(\ref{mu2propa}), we deduce that the source term derives from an interaction
Lagrangian of the form
\begin{equation}\label{Sint}
 S_{\rm int} = \int\dd\eta\dd^3\bx \frac{4a}{\sqrt{8\pi G}}\left[\partial_i\Phi\partial_j\Phi
 +4\pi G\,\partial_i\chi\partial_j\chi\right]
 \mu^{ij}.
\end{equation}
It describes a two-scalars graviton interaction. In full generality the interaction term would
also include, at lowest order, cubic terms of three scalars, two gravitons-scalar and three
gravitons. They respectively correspond to second order scalar-scalar modes generated from
gravitational waves and second order tensor modes. As emphasized previously, we do not consider
these interactions here.

%%%%%%%%%%%%%%%%%%%%%%%%%%%%%%%%%%%
\section{Comparison of the two formalisms}\label{sec4b}
%%%%%%%%%%%%%%%%%%%%%%%%%%%%%%%%%%%

Before going further it is instructive to compare the two formalisms and understand how they relate to each other. 
Note that we go beyond Ref.~\cite{bi:BDE}, where a comparison of the variables was made at linear 
order. Here we investigate how the equations map to each other and extend the discussion to second order for the tensor sector. At the background level the scale factors $a$ and expansion rates $H$ introduced in each formalism agree, which explains why we made use of the same notation. 

The perturbations of the metric around FL space-time has been split into a first-order and a
second-order part according to
 \be\label{decomposition-ordre2}
  X=X^{(1)}+\frac{1}{2}X^{(2)}\,.
 \ee
We make a similar decomposition for the quantities used in the $1+3$ covariant formalism. As long
as we are interested in the gravitational wave sector, we only need to consider the four-velocity
of the perfect fluid describing the matter content of the universe which we decompose as
\begin{equation}
 u^{\mu}=\frac{1}{a}(\delta_{0}^{\mu} + V^{\mu}).
\label{defvelocity}
\end{equation}
Its spatial components are decomposed as
\begin{equation}
 V^{i}=\partial^{i}V + \bar{V}^{i}\,,
\end{equation}
$\bar{V}^{i}$ being the vector degree of freedom and $V$ the scalar degree of freedom. As $V^\mu$ has only three independent degrees of freedom since $u^{\mu}$ satisfies
$u_{\mu}u^{\mu}=-1$, its temporal component is linked to other perturbation variables. We assume
that the fluid has no vorticity ($\bar{V}^{i}=0$), as it is the case for the scalar fluid we have in mind and consequently we will also drop the vectorial perturbations ($\bar{\mathcal E}_{i}=0$).

%%%%%%%%%%%%%%%%%%%%%%%%%%%%%%%%%%%%%%%%%%%%
\subsection{Matching at linear order}
%%%%%%%%%%%%%%%%%%%%%%%%%%%%%%%%%%%%%%%%%%%%

At first order, the spatial components of the shear, acceleration and expansion are respectively
given by
\begin{equation}\label{geo1}
  \sigma_{ij}^{(1)} = a \left(\p_{\langle i}\p_{j \rangle} V^{(1)}  + \bar{\mathcal{E}}^{{(1)}\prime}_{ij}
  \right)\,,
\end{equation}
\begin{equation}\label{geo2}
  \acc_{i}^{(1)} = \p_{i} \left( \Phi^{(1)} + {\cal{H}} V^{(1)} + V^{(1)\prime} \right)\,,
\end{equation}
\begin{equation}\label{geo3}
  {\delta \Theta}^{(1)} = \frac{1}{a}\left( -3 \Psi^{(1)\prime} - 3{\cal{H}} \Phi^{(1)}
  + \Delta V^{(1)} \right)\,.
\end{equation}
The electric and magnetic part of the Weyl
tensor take the form
\begin{equation}\label{geo4}
  E_{ij}^{(1)} = \p_{\langle i}\p_{j \rangle} \Phi^{(1)} - \frac{1}{2}\left(
  \bar{\mathcal E}''_{ij} + \Delta \bar{\mathcal E}_{ij}\right)\,,
\end{equation}
\begin{equation}\label{geo5}
  H_{ij}^{(1)} = \eta_{kl\langle i}\p^{k}\bar{\mathcal E}^{(1)\prime\,\,\,l}_{j\rangle}
  \equiv (\hat{\curl}\bar{\mathcal E}^{(1)\prime})_{ij}\,.
\end{equation}
Note that $\eta_{kli}$ is the completely antisymmetric tensor normalized such that $\eta_{123}=1$,
which differs from $\varepsilon_{abc}$. We deduce from the last expression
that
\begin{equation}\label{geo6}
  \left(\curl E^{(1)}\right)_{ij}=
  -\frac{1}{2a} \left[\left(\hat{\curl} \bar{\mathcal E}^{(1)\prime\prime}\right)_{ij}
  + \left(\hat{\curl} \Delta\bar{\mathcal E}^{(1)}\right)_{ij}
 \right]\;,
 \end{equation}
where we have used simpler notation by  writing $(\hat{\curl}\bar{\mathcal E})_{ij}$ as $\hat{\curl}
\bar{\mathcal E}_{ij}$.
We also note that the derivative along $u_{\mu}$ of a tensor $T$ of rank $(n,m)$, vanishing in the
background, takes the form
\begin{equation}\label{eq:dercomp1}
 \dot{T}^{i_1 ... i_n}_{\,\,\,\,j_1 ...j_m} =
 \partial_t T^{i_1 ... i_n}_{\,\,\,\,j_1 ... j_m}+(n-m)H T^{i_1 ... i_n}_{\,\,\,\,j_1
... j_m}
\end{equation}
at first order, or alternatively
\begin{equation}\label{eq:mapping-dot}
\frac{\left(a^{m-n}T^{i_1 ... i_n}_{\,\,\,\,j_1 ...j_m}\right)^{.}}{a^{m-n}} = \partial_tT^{i_1
... i_n}_{\,\,\,\,j_1 ... j_m}\,.
\end{equation}
Again, recall that a dot refers to a derivative along $u^\mu$. Indeed at first order, it reduces
to a derivative with respect to the cosmic time but this does not generalize to second order.

Now, Eq.~(\ref{eq:dotH}) can be recast a
\begin{equation}\label{eq:dotH2}
 a^{-2}\left(a^2 H_{ij}\right)^{.} + \curl E_{ij} + H\, H_{ij}=0\,.
\end{equation}
Using the expressions (\ref{geo4}-\ref{geo5}) for the geometric quantities, this equation takes
the form
\begin{equation}
 \hat\curl\left[\frac{1}{2a}\left( \bar {\mathcal E}_{ij}'' + 2{\cal{H}}\bar
 {\mathcal E}_{ij}'-\Delta\bar {\mathcal E}_{ij}\right)\right]=0\,.
\end{equation}
Similarly Eq.~(\ref{eq:dotdotH}) can be recast as
\begin{equation}\label{eq:dotdotH1}
 \frac{\left(a^2 {H}_{ab}\right)^{..}}{a^2} +3 H \frac{\left(a^2 {H}_{ab}\right)^{.}}{a^2}
 + 2(H^2+\dot{H})H_{ab} -\n^{2}H_{ab}= 0\,,
\end{equation}
so that it reduces at first order to
\begin{equation}\label{e112}
 \hat\curl\left[\frac{1}{2 a^2}\left( \bar{\mathcal E}_{ij}^{(1)\prime\prime} +
     2{\cal{H}}\bar {\mathcal E}_{ij}^{(1)\prime}-
     \Delta\bar{\mathcal E}^{(1)}_{ij}\right)'\right]=0\,.
\end{equation}

Thus, Eq.~(\ref{e112}) maps to Eq.~(\ref{Ebar1}) with the
identification~(\ref{geo5}), if there is
no vector modes. This can be understood from the fact that in the Bardeen formalism,
Eq.~(\ref{Ebar1}) is obtained from the Einstein equation as $\hat\curl^{-1}[\hat\curl G_{ij}]=0$.

In the case where there are vector modes, Eq.~(\ref{geo5}) has to be replaced by
$$
 H_{ij}^{(1)} =  (\hat{\curl}\bar{\mathcal E}^{(1)\prime})_{ij} + 
 \frac{1}{2}\eta_{kl\langle i}\p^{k} \p_{j \rangle}\bar{\mathcal E}^{(1)\prime\,l}\,
$$
and $H_{ab}$ is no longer a description of the GW, i.e. directly related to the
TT part of the spacetime metric and the matching is not valid anymore.

%%%%%%%%%%%%%%%%%%%%%%%%%%%%%%%%%%%%%%%%%%
\subsection{Matching at second order}
%%%%%%%%%%%%%%%%%%%%%%%%%%%%%%%%%%%%%%%%%%

At second order, the matching is much more intricate mainly because
the derivative along $u^\mu$ does not match with the derivative respect to cosmic time any more.

Let us introduce the short hand notation
\begin{equation}
\left(X \times Y \right)_{ij} \equiv \eta_{kl\langle i} X^{k}Y^{\,\,\,l}_{j\rangle}
\end{equation}
for any tensors $X^k$ and $Y^{lm}$. If $Y^{lm}=\p^l \p^m Z$, or $X^k=\p^k W$, we also use the
short-hand notation $Y=\p\p Z$ $X=\p W$.

Among the terms quadratic in first-order perturbations, those involving a first-order tensorial
perturbation can be omitted, as we are only interested in second-order effects sourced by scalar
contributions. At second order, the geometric quantities of interest read
\begin{widetext}
\begin{eqnarray}
 H_{ij}^{(2)}&=& \left(\hat{\curl}\bar{\mathcal E}^{(2)\prime}\right)_{ij}-4 \left(\p V^{(1)}\times \p \p
                  \Phi^{(1)}\right)_{ij}\\
 \left(\curl E^{(2)}\right)_{ij}&=&
  -\frac{1}{2a} \left[\left(\hat{\curl} \bar{\mathcal E}^{(2)\prime\prime}\right)_{ij}
                + \left(\hat{\curl} \Delta\bar{\mathcal E}^{(2)}\right)_{ij}\right]
      - \frac{2}{a} \left[\left(\p \Phi^{(1)} \times \p \p \Phi^{(1)} \right)_{ij}
         + {\cal{H}}\left(\p V^{(1)} \times \p \p \Phi^{(1)} \right)_{ij}\right. \nonumber\\
&& \qquad \qquad \qquad \qquad \qquad \qquad \qquad \qquad \qquad \qquad\left. - \left(\p V^{(1)} \times \p \p \Phi^{(1)\prime} \right)_{ij} \right]\,.
\end{eqnarray}
\end{widetext}
From the latter expression, we remark that $H^{(2)}_{ij}$ has a term quadratic in first-order
perturbations involving $V^{(1)}$ and $\Phi^{(1)}$. This terms arise from a difference between the
two formalisms related to the fact that geometric quantities, such as $H_{ij}$ $E_{ij}$ etc., live
on the physical space-time, whereas in perturbation theory, any perturbation variable at any
order, such as $V^{(1)}$, ${\mathcal E}^{(2)}_{ij}$ etc., are fields propagating on the background
 space-time.

It follows that the splitting into tensor, vector and scalar modes is different. In the covariant
formalism, the splitting refers to the fluid on the physical space-time, whereas in perturbation
theory it refers to the co-moving fluid of the background solution. Indeed, this difference only
shows up at second order as the magnetic Weyl tensor vanishes in the background. The one to one
correspondence at first order between equations of both formalisms disappears, as the second order
equations of the covariant formalism contain the dynamics of the first order quantities.

When keeping terms contributing to the second order, Eq.~(\ref{eq:dotH}) has an additionnal source
term and reads
\begin{equation}
\dot H_{ab}+\curl E_{ab} + 3H H_{ab} = -2 \epsilon_{cd\langle a} \acc^{c} E^{\,\,\,d}_{b\rangle}
\label{eq:dotH2bis}
\end{equation}
If first order  tensorial perturbations are neglected then $H_{ab}$ vanishes at first order and
Eq.~(\ref{eq:dercomp1}) still holds when applied to $H_{ab}$. Thus Eq.~(\ref{eq:dotH2bis}) can be
recast as
\begin{equation}
\frac{\left(a^2 H_{ab}\right)^{.}}{a^2} + \curl E_{ab} + \frac{\cal{H}}{a} H_{ab} = -2
\epsilon_{cd\langle a} \acc^{c} E^{\,\,\,d}_{b\rangle} \;.\label{eq:dotH3}
\end{equation}

Substituting the geometric quantities for their expressions at second order, and making use of
Eq.~(\ref{eq:mapping-dot}) to handle the derivatives, Eq.~(\ref{eq:dotH2}) reads at second order
\begin{widetext}
\begin{equation}
\hat{\curl}\left[\frac{1}{2 a}\left(  \bar{\mathcal E}_{ij}^{(2)\prime\prime} +
2{\cal{H}}\bar{\mathcal E}_{ij}^{(2)\prime}-\Delta\bar{\mathcal E}_{ij}^{(2)}\right)\right]=
-\frac{2}{a} \left[ \left( \p \Phi^{(1)} \times \p \p \Phi^{(1)} \right)_{ij} - \left( \p V^{(1)}
\times \p \p \Phi^{(1)\prime} \right)_{ij} - {\cal{H}} \left( \p V^{(1)} \times \p \p \Phi^{(1)}
\right)_{ij} \right]\;.
\end{equation}
%\end{widetext}
Using the momentum and constraint equation (\ref{e41}) at first order
\begin{equation}
\Phi^{(1)\prime} + {\cal{H}}\Phi^{(1)}=\left({\cal{H}}' - {\cal{H}}^2 \right) V^{(1)}
\end{equation}
and the background equation ${\cal{H}}' - {\cal{H}}^2 = -4 \pi G\mu(1+w) a^2$, that we deduce
from the Raychaudhuri equation and the Gauss-Codacci equation at first order, we can link it to Eq.~(\ref{Ebar2})
as it then reads
%\begin{widetext}
\begin{equation}
\frac{1}{a}\hat{\curl}\left[\frac{1}{2}\left(  \bar{\mathcal E}_{ij}^{(2)\prime\prime} +
2{\cal{H}}\bar{\mathcal E}_{ij}^{(2)\prime}-\Delta\bar{\mathcal E}_{ij}^{(2)}\right)\right]=
\frac{1}{a} \hat{\curl}\left[ 2 \p_i \Phi^{(1)} \p_j \Phi^{(1)} + 8 \pi G a^2 (\mu + P) \p_i
V^{(1)} \p_j V^{(1)} \right]\;.
\end{equation}
When applied to a scalar field, this is exactly the gravitational wave propagation
equation~(\ref{Ebar2}) with the source term~(\ref{sourceterm}). 
%Exactly as at linear order, the
%mapping is made up to a curl-free integration constant.
\end{widetext}

%%%%%%%%%%%%%%%%%%%%%%%%%%%%%%%%%%%%%%%%%%%%%
\subsection{Discussion}
%%%%%%%%%%%%%%%%%%%%%%%%%%%%%%%%%%%%%%%%%%%%%

In conclusion, we have matched both the perturbation variables and equations at first and second
order in the perturbations. This extends the work of Ref.~\cite{bi:BDE} which considered the linear case, and has not been previously investigated.

Even though we restrict to the tensor sector, this comparison is instructive and illustrates the
difference of approach between the two formalisms, in a clearer way than at first order. In the
Bardeen approach, all perturbation variables live on the unperturbed spacetime. At each order, we
write exact equations for an approximate spacetime. In particular, this implies that the time
derivatives are derivative with respect to the cosmic time of the background spacetime. In the
covariant approach, one derives an exact set of equations (assuming no perturbation to start
with). These exact equations are then solved iteratively starting from a background solution which
assumes some symmetries. The time derivative is defined in terms of the flow vector as
$u^a\nabla_a$. Indeed, at first order for scalars, this derivative matches exactly with the
derivative with respect to the background cosmic time. At second order, this is no longer the
case. First the flow vector at first order does not coincide with its background value. This
implies a (first-order) difference between the two time derivatives which must be taken into
account. Then, the geometric quantities, such as $H_{ij}$ $E_{ij}$ etc., ``live" on the physical
space-time, whereas in perturbation theory, any perturbation variable at any order, such as
$V^{(1)}$, ${\mathcal E}^{(2)}_{ij}$ etc., live on the background space-time. This explains why
e.g. $H^{(2)}_{ij}$ has a term quadratic in first-order perturbations involving $V^{(1)}$ and
$\Phi^{(1)}$.

The master variables and corresponding wave equations in both formalisms are
also different in nature. In the metric approach the wave equation with source
is defined non-locally in Fourier space; in the covariant approach, we are
able to derive a local tensorial wave equation which, because it is
divergence-free, represents the gravitational wave contribution. Of course, we
can make a non-local decomposition in Fourier space as required. Furthermore,
on one hand the TT part of the metric in a particular gauge is a perturbative
approach used to describe GW, and this tells us the shear of spatial lengths
with respect to a homogenous and isotropic background, referring implicitly to
a hypothetical set of averaged observers. On the other hand, the covariant
description using $H_{ab}$ which is built out of the Weyl tensor and the
comoving observer's velocity, directly describes the dynamically free part of
the gravitational field~\cite{royetal} (up to second-order when rotation is
zero) as seen by the true comoving observers. This is part of the dynamic
spacetime curvature which  directly induces the motion of test particles
through the geodesic deviation equation, and it accounts for effects due to the non-homogenous comoving fluid velocity.  

There is one more difference between the two formalisms, concerning the initial conditions. In
the Bardeen approach, as we recalled in section~\ref{sec4}, there is a natural way to set up the
initial conditions on sub-Hubble scales by identifying canonical variables, both for the scalar and
tensor modes, and promoting them to the status of quantum operators. In the covariant formalism
such variables have not been constructed in full generality (see however Ref.~\cite{pu06} for a
proposal). Consequently this sets limitations to this formalism since it cannot account for both the evolution and the initial conditions at the same time.

%%%%%%%%%%%%%%%%%%%%%%%%%%%%%%%%%%
\section{Illustration: slow-roll inflation}\label{sec5}
%%%%%%%%%%%%%%%%%%%%%%%%%%%%%%%%%%

%%%%%%%%%%%%%%%%%%%%%%%%%%%%%%%%%%.
\subsection{Slow-roll inflation}
%%%%%%%%%%%%%%%%%%%%%%%%%%%%%%%%%%

In this section, we focus on the case of a single slow-rolling scalar field and we introduce the
slow-roll parameters
\begin{equation}\label{eq:IV-1}
 \varepsilon = \frac{3}{2}\frac{\psi^2}{\mu},\qquad
 \delta = -\frac{\dot\psi}{H\psi}.
\end{equation}
Using the Friedmann equations~(\ref{eq:e1b}-\ref{eq:e1a}), these parameters can be expressed in
terms of the Hubble parameter as
\begin{equation}\label{eq:IV-2}
 \varepsilon = -\frac{1}{4\pi G}\left[\frac{H'(\phi)}{H(\phi)}\right]^2,\qquad
 \delta = \frac{1}{4\pi G}\frac{H''(\phi)}{H(\phi)}.
\end{equation}
Interestingly Eq.~(\ref{eq:e1b}) takes the form
\begin{equation}
 H^2\left(1-\frac{1}{3}\,\varepsilon\right) = \frac{\kappa}{2}V(\phi),
\end{equation}
which implies
\begin{equation}
 \frac{\ddot a}{a} = (1-\varepsilon)H^2.
\end{equation}
The equation of state and the sound speed of the equivalent scalar field are thus given by
\begin{equation}
 w=-1+\frac{2}{3}\varepsilon,\qquad
 c_s^2=-1+\frac{2}{3}\delta.
\end{equation}
The evolution equations for  $\varepsilon$ and $\delta$ show that $\dot\varepsilon$ and
$\dot\delta$ are of order 2 in the slow-roll parameters so that at first order in the slow-roll
parameters, they can be considered constant. Using the definition of the conformal time and
integrating it by parts, one gets
\begin{equation}\label{aH}
 a(\eta) = -\frac{1}{H\eta}\frac{1}{1-\varepsilon},
\end{equation}
assuming $\varepsilon$ is constant, from which it follows that
\begin{equation}
 \mathcal{H}\equiv aH = -\frac{1}{\eta}(1+\varepsilon) +\mathcal{O}(2),
\end{equation}
where $\eta$ varies between $-\infty$ and $0$. This implies that
\begin{equation}
 \frac{a''}{a}=\frac{2+3\varepsilon}{\eta^2},\qquad
 \frac{z''}{z}=\frac{2+6\varepsilon-3\delta}{\eta^2}.
\end{equation}
The general solution of Eq.~(\ref{kgv}) is 
\begin{equation}
v_k=\sqrt{-\pi\eta/4}\left[c_1H_\nu^{(1)}(-k\eta)+c_2
H_\nu^{(2)}(-k\eta)\right]\;,
\end{equation}
with $|c_1|^2-|c_2|^2=1$, where $H_\nu^{(1)}$ and $H_\nu^{(2)}$ are
Hankel functions of first and second kind and $\nu=3/2 + 2\varepsilon-\delta$. Among
this family of solutions, it is natural to choose the one with $c_2=0$ which contains only
positive frequencies~\cite{bmf}. It follows that the solution with these initial conditions is
\begin{equation}
 v_k(\eta)=\frac{\sqrt{\pi}}{2}\sqrt{-\eta}H_\nu^{(1)}(-k\eta)\,.
\label{eq:evolution_vk}
\end{equation}

On super-Hubble scales, $|k\eta|\ll1$, we have
$$
v_k\rightarrow 2^{\nu-3/2}\Gamma(\nu)/\Gamma(3/2)(2k)^{-1/2} (-k\eta)^{-\nu+1/2}.
$$
Now, using Eq.~(\ref{aH}) to express $\eta$ and Eq.~(\ref{za}) to replace $z$ in
expression~(\ref{PR}), we find that
\begin{eqnarray}
 {\cal P}_{\mathcal{R}}(k) &=& \frac{1}{\pi}\,\frac{H^2}{M_p^2\varepsilon}
 \left[2^{\nu-3/2}\frac{\Gamma(\nu)}{\Gamma(3/2)}\right]^2\left(\nu-\frac{1}{2}\right)^{-2\nu+1}
 \nonumber\\
 &&\qquad\qquad\qquad\times
 \left(\frac{k}{aH}\right)^{-2\nu+3}\;,
\end{eqnarray}
where we have set $M_p^2=G^{-1}$. At lowest order in the slow-roll parameter, it reduces to
\begin{eqnarray}
 {\cal P}_{\mathcal{R}}(k) &=& \frac{1}{\pi}\,\frac{H^2}{M_p^2\varepsilon}
 \left(\frac{k}{aH}\right)^{2\delta-4\varepsilon}.
\label{PR1}
\end{eqnarray}

The evolution of the gravitational waves at linear order are dictated by the same equation but
with $\nu_T=3/2 + \varepsilon$, so that
\begin{equation}
 \mu^{(1)}_k(\eta)=\frac{\sqrt{\pi}}{2}\sqrt{-\eta}H_{\nu_T}^{(1)}(-k\eta).
\end{equation}
Similarly as for the scalar mode, we obtain
\begin{eqnarray}
{\cal P}_{T}(k) &=& \frac{16}{\pi}\,\frac{H^2}{M_p^2}
 \left(\frac{k}{aH}\right)^{-2\varepsilon}.
\end{eqnarray}

%%%%%%%%%%%%%%%%%%%%%%%%%%%%%%%%%%%%%%%%%
\subsection{Gravitational waves at second order}
%%%%%%%%%%%%%%%%%%%%%%%%%%%%%%%%%%%%%%%%%

The couplings between scalar and tensor modes at second order imply that the second order
variables can be expanded as
$$
 \mathcal{R}=\mathcal{R}^{(1)}+\frac{1}{2}\left(\mathcal{R}^{(2)}_{\mathcal{R}\mathcal{R}}+
 \mathcal{R}^{(2)}_{{\mathcal E}{\mathcal E}}+\mathcal{R}^{(2)}_{\mathcal{R}{\mathcal E}}\right)
$$
and a similar expansion for ${\mathcal E}$, where, e.g., $\mathcal{R}^{(2)}_{\mathcal{R}{\mathcal E}}$ stands for the
second order scalar modes induced by the coupling of first order scalar and tensor modes etc. The
deviation from Gaussianity at the time $\eta$ of the end of inflation can be characterized by a series of coefficients $f_{\rm NL}^{a,bc}$
defined for example as
\begin{eqnarray}
\frac{1}{2}\mathcal{R}^{(2)}_{{\mathcal E}{\mathcal E}}(\bk,\eta)&=& \frac{1}{(2 \pi)^{3/2}}\int\delta^3(\bk_1+\bk_2-\bk)
 {\mathcal E}(\bk_1,\eta){\mathcal E}(\bk_2,\eta)\nonumber\\
 &&\qquad\quad f^{\mathcal{R},{\mathcal E}{\mathcal E}}_{\rm NL}(\bk,\bk_1,\bk_2,\eta)\dd^3\bk_1\dd^3\bk_2.
\end{eqnarray}
These six coefficients appear in different combinations in the connected part of the 3-point
correlation function of $\mathcal{R}$ and ${\mathcal E}$. For instance
\begin{widetext}
\begin{equation}
 \langle {\mathcal E}_{\bk_1}{\mathcal{R}}_{\bk_2}{\mathcal{R}}_{\bk_3}\rangle_c =
   \left[2 f^{{\mathcal E},\mathcal{R}\mathcal{R}}_{\rm NL}(\bk_1,\bk_2,\bk_3)
   P_{\mathcal{R}}(k_3)P_{\mathcal{R}}(k_2) + f^{{\mathcal R},\mathcal{E}\mathcal{R}}_{\rm NL}(\bk_1,\bk_2,\bk_3)
   P_{\mathcal{R}}(k_3)P_{\mathcal{E}}(k_1) \right] \delta^3(\bk_1+\bk_2+\bk_3)
\label{eq:ERR}
\end{equation}
\end{widetext}
and $f^{\mathcal{R},\mathcal{R}\mathcal{R}}_{\rm NL}$ is the standard $f_{\rm NL}$ parameter. One
can easily check that $\langle {\mathcal
R}_{\bk_1}\mathcal{R}_{\bk_2}\mathcal{R}_{\bk_3}\rangle_c$ involves $f^{{\mathcal
R},\mathcal{R}\mathcal{R}}_{\rm NL}$, $\langle {\mathcal E}_{\bk_1}{\mathcal
E}_{\bk_2}\mathcal{R}_{\bk_3} \rangle_c$  involves $f^{{\mathcal E},{\mathcal E}\mathcal{R}}_{\rm
NL}$ and $f^{\mathcal{R},{\mathcal E}{\mathcal E}}_{\rm NL}$, and
$\langle{\mathcal E}_{\bk_1}{\mathcal E}_{\bk_2}{\mathcal E}_{\bk_3}\rangle_c$ involves $f^{{\mathcal E},{\mathcal E}{\mathcal E}}_{\rm NL}$.\\

%%%%%%%%%%%%%%%%%%%%%%%%%%%%%%%%%%%%%%%%%%%%.
\subsection{Expression for $f^{{\mathcal E},\mathcal{R}\mathcal{R}}_{\rm NL}$}
%%%%%%%%%%%%%%%%%%%%%%%%%%%%%%%%%%%%%%%%%%%%

From our analysis, we can give the expression of $f^{{\mathcal E},\mathcal{R}\mathcal{R}}_{\rm NL}$.
Starting from the fact that $-\varepsilon\mathcal{R}=\Phi(1+\varepsilon)+\Phi'/\mathcal{H}$ and
from the expression~(\ref{chiQ}), we get that $\Phi\sim-\varepsilon\mathcal{R}-\Phi'/\mathcal{H}$
or $\Phi=-\epsilon \eta \int \frac{\mathcal{R}}{\eta^{2}} \dd\eta $, and $\sqrt{4\pi G}\chi\sim
-\sqrt{\varepsilon}\left[\mathcal{R} - \Phi'/\mathcal{H}\right]$. It follows that the source
term~(\ref{sourceterm}) reduces at lowest order in the slow-roll parameter to
$$
 S_{ij}^{\rm TT}=4\left[\varepsilon\partial_i\mathcal{R}\partial_j\mathcal{R} \right]^{\rm TT}.
$$
The interaction Lagrangian is thus given by
\begin{equation}\label{Sint2}
 S_{\rm int} = \int\dd\eta\dd^3\bx \frac{4a}{\sqrt{8\pi G}}\,
 \varepsilon\,\partial_i\mathcal{R}\partial_j\mathcal{R}\,
 \mu^{ij},
\end{equation}
which reduces to
\begin{equation}\label{Sint3}
 S_{\rm int} = \int\dd\eta\dd^3\bx
 \, 2 \partial_i v \, \partial_j v\,
  \bar{\mathcal E}^{ij}.
\end{equation}
This is the same expression as obtained in Ref.~\cite{maldacena}.

In full generality, during inflation, we should use the ``in-in'' formalism to compute any
correlation function of the interacting fields. As was shown explicitly in Ref.~\cite{bbu} for a
self-interacting field and more generally in Ref.~\cite{weinberg}, the quantum computation agrees
with the classical one on super-Hubble scales at lowest order. Note however that both computations
may differ (see Ref.~\cite{Acq} versus Ref.~\cite{maldacena}) due to the fact that in the
classical approach the change in vacuum is ignored. The difference does not affect the order of
magnitude but the geometric $k$-dependence. In order to get an order of magnitude, we thus
restrict our analysis here to the classical description. This description is also valid when
considering the post-inflationary era.

In the classical approach, we can solve Eq.~(\ref{eqmu2}) by mean of a Green function. Since the two
independent solutions of the homogeneous equation are $\sqrt{-k
  \eta}H_{\nu_T}^{(1/2)}(-k \eta)$, the Wronskian
of which is $4i/(\pi k)$, the Green function is given by
\begin{eqnarray}
 G(k,\eta,\eta') &=& -i\frac{\pi}{4}\sqrt{\eta\eta'}\left[
 H_{\nu_T}^{(1)}(-k\eta) H_{\nu_T}^{(2)}(-k\eta')\right.\nonumber\\
 &&\qquad\left.- H_{\nu_T}^{(1)}(-k\eta') H_{\nu_T}^{(2)}(-k\eta)
 \right].
\end{eqnarray}
It follows that the expression of the second order tensor perturbation is given by
\begin{eqnarray}
 \mu^{(2)}_{\bk,\lambda}(\eta)&=&\frac{2}{(2 \pi)^{3/2}}\int_{-\infty}^{\eta}\dd\eta' \frac{a(\eta')}{\sqrt{8\pi G}}\,\varepsilon\,G(k,\eta,\eta')
 \nonumber\\
 &&\quad\int \dd^3\bq \left(q_iq_j\varepsilon^{ij}_\lambda\right)
 \mathcal{R}_{\bq}(\eta')\mathcal{R}_{\bk-\bq}(\eta').
\end{eqnarray}
We thus obtain
\begin{widetext}
\begin{eqnarray}
f^{{\mathcal E}_{\lambda}  \mathcal{R}
  \mathcal{R}}_{NL}(\bk,{\bm{q_1}},{\bm{q_2}},\eta) = \left[ \mathcal{R}_{\bm{q_1}}(\eta)\mathcal{R}_{\bm{q_2}}(\eta) \right]^{-1} \frac{\varepsilon}{a(\eta)} \int_{-\infty}^{\eta}\dd\eta'
a(\eta')\,G(k,\eta,\eta')
\left(q_{1i}q_{1j}\varepsilon^{ij}_\lambda(\bk)\right)
 \mathcal{R}_{\bm{q_1}}(\eta')\mathcal{R}_{\bm{q_2}}(\eta')\;.
\end{eqnarray}
\end{widetext}

If we want to estimate Eq.~(\ref{eq:ERR}) in the squeezed limit  $k_1 \ll k_2,k_3$ the
contribution coming from the term involving $f^{{\mathcal E}_{\lambda}  \mathcal{R}
  \mathcal{R}}_{NL}(\bk,{\bm{q_1}},{\bm{q_2}},\eta)$ can be computed by use
of the super-Hubble limit of the Green function $|G(k, \eta ,\eta')| \simeq
\frac{\sqrt{\eta \eta'}}{2 \nu_T}\left[
  \left(\frac{\eta'}{\eta}\right)^{\nu_T}-\left(\frac{\eta'}{\eta}\right)^{-\nu_T}\right]$.
This contribution will be proportional to $\frac{H^4}{M_p^4 \varepsilon}
k_2^{-8}\left(k_{2i}k_{2j}\varepsilon^{ij}_\lambda\right)\delta^3(\bk_1+\bk_2+\bk_3)$, which is
the same order of magnitude as in Ref.~\cite{maldacena}, but do not have the same geometric
dependence as it goes like $k_2^{-5} k_1^{-3}$ instead.

%%%%%%%%%%%%%%%%%%%%%%%%%%%%%%%%%%%
\subsection{Orders of magnitude}
%%%%%%%%%%%%%%%%%%%%%%%%%%%%%%%%%%%

When we want to estimate $\langle{\mathcal E}^{(2)}_{\bk,\lambda}(\eta){\mathcal
E}^{(2)*}_{\bk',\lambda'}(\eta)\rangle$, we have to evaluate the connected part of $\langle
\mathcal{R}_{\bq}(\eta')\mathcal{R}_{\bk-\bq}(\eta')
\mathcal{R}^*_{\bp}(\eta'')\mathcal{R}^*_{\bk'-\bp}(\eta'')\rangle$, where $\bq$ and $\bp$ are the
two internal momentum and $\eta'$ and $\eta''$ the two integration times. From the Wick theorem,
this correlator reduces to $\mathcal{R}(q,\eta')\mathcal{R}^*(q,\eta'')
\mathcal{R}(|\bk-\bq|,\eta')\mathcal{R}^*(|\bk-\bq|,\eta'')\delta(\bk-\bk')[\delta(\bq-\bp)+
\delta(\bk-\bq-\bp)]$ and because $k^i\varepsilon_{ij}=0$ the two terms give the same geometric
factor. Thus, the integration on $\bp$ is easily done and we can factorize $\delta(\bk-\bk')$.
Now, note that the terms in the integral involve only the modulus of $\bq$ and $\bk-\bq$ so that
it does not depend on the angle $\varphi$ of $\bq$ in the plane orthogonal to $\bk$. This implies
that the integration of $\varphi$ will act on a term of $\cos^22\varphi$, $\sin^22\varphi$ and
$\cos2\varphi\sin2\varphi$ respectively for $++$, $\times\times$ and $+\times$ so that it gives a
term $\pi\delta_{\lambda\lambda'}$. In conclusion, defining the second order power spectrum ${\cal
P}_T^{(2)}$ by
\begin{equation}\label{specE2}
 \frac{1}{4}\langle {\mathcal E}^{(2)}_{\bk,\lambda} {\mathcal E}^{*{2}}_{\bk',\lambda'}\rangle = \frac{2\pi^2}{k^3}{\cal P}^{(2)}_T(k)
 \delta^{(3)}(\bk-\bk')\delta_{\lambda\lambda'},
\end{equation}
it can be expressed as
\begin{widetext}
\begin{eqnarray}
 {\cal P}^{(2)}_T(k)&=& \frac{k^3}{(2 \pi)^3 \pi^{2} a^2}\int\dd\eta'\dd\eta'' a(\eta')a(\eta'')\,\varepsilon^2\,G(k,\eta,\eta')
 G^*(k,\eta,\eta'')\nonumber\\
&&\times \int\dd^3q\left(q_iq_j\varepsilon^{ij}_\lambda\right)^2\mathcal{R}(q,\eta')\mathcal{R}^*(q,\eta'')
\mathcal{R}(|\bk-\bq|,\eta')\mathcal{R}^*(|\bk-\bq|,\eta'').
\end{eqnarray}
%\end{widetext}
Setting $\bk\cdot\bq=kq\mu$, this reduces to
%\begin{widetext}
\begin{eqnarray}
 {\cal P}^{(2)}_T(k)&=&\frac{ k^3}{(2 \pi)^3 \pi a^2} \int\,q^6\dd q \left(1-\mu^2\right)^2\dd\mu
 \left|\int_{-\infty}^{\eta}\,\dd\eta'\,a(\eta')\,\varepsilon\,G(k,\eta,\eta')\mathcal{R}(q,\eta')
\mathcal{R}(|\bk-\bq|,\eta')\right|^2,
\end{eqnarray}
after integration over $\varphi$ which gives a factor $\pi\left(1-\mu^2\right)^2q^4$.
\end{widetext}
We can now take the super-Hubble limit of this expression at lowest order in
the slow-roll parameters. In order to do so, we make use of the super-Hubble
limit of the Green function given above, and we perform the time integral from $1/k$ to $\eta$ and keep only the leading order contribution:
\begin{eqnarray}
 {\cal P}^{(2)}_T(k)&=&\frac{1}{3^4 2^3 \pi^2}  G^2 H^4 F(\epsilon,\delta) \left(\frac{k}{aH}\right)^{-2 \epsilon}\;,
\end{eqnarray}
where, with the defnitions ${\bm{y}} \equiv \bq/k$ and ${\bm{n}} \equiv\bk/k$, 
\begin{equation}
F(\epsilon,\delta) \equiv \int\, \left(y \left|{\bm{n}}-{\bm{y}}\right|\right)^{-3 -4\epsilon + 2 \delta} y^6\dd y \left(1-\mu^2\right)^2\dd\mu
\end{equation}
is a numerical factor.
In this approximation, the ratio between the second order power spectrum and the first order power spectrum at leading order in the slow-roll parameters, is given by:
\begin{equation}
 \frac{{\cal P}^{(2)}_T(k)}{{\cal P}^{(1)}_T(k)} = \frac{1}{2^7 3^4 \pi}\left(\frac{H}{M_p}\right)^{2} F(\epsilon,\delta).
\end{equation}

Indeed there are ultraviolet and infrared divergences hidden in $F(\epsilon,\delta)$. We expect
the infrared divergence not to be relevant for observable quantities due to finite volume effects
(see for instance Ref.~\cite{finite}). The ultraviolet divergence, on the other hand, has to be
carefully dimensionaly regularized in the context of quantum field theory (see e.g.
Ref.~\cite{weinberg}).

\vskip3cm

%%%%%%%%%%%%%%%%%%%%%%%%%%%%%%%%%%%%%%%%%%%%%%%%%%%%%%%%%%%%%%%%%%%%%%%%%%%%%%%%%%%%%%%%%
\section{Conclusions}
%%%%%%%%%%%%%%%%%%%%%%%%%%%%%%%%%%%%%%%%%%%%%%%%%%%%%%%%%%%%%%%%%%%%%%%%%%%%%%%%%%%%%%%%%

In this article we have investigated the generation of gravitational waves due to second order
effects during inflation.
We have considered these effects both in the covariant perturbation formalism and in the more
standard metric based approach. The relation between the two formalisms at second-order has been
considered and we have discussed their relative advantages. This comparison leads to a better
understanding of the differences in dynamics between the two formalisms.

As an illustration, we have focused on GW generated by the coupling of first order scalar modes.
To characterize this coupling we have introduced and computed the parameter $f^{{\mathcal
E},\mathcal{R}\mathcal{R}}_{\rm NL}$. It enters in the expression of $\langle {\mathcal
E}_{\bk_1}{\mathcal R}_{\bk_2}\mathcal{R}_{\bk_3} \rangle_c$ that was shown to be of order
$(H/M_p)^4/\varepsilon$, as $\langle {\mathcal R}_{\bk_1}{\mathcal R}_{\bk_2}\mathcal{R}_{\bk_3}
\rangle_c$. On the other hand the power spectrum of GW remains negligible.

This shows that the contribution of $\langle {\mathcal E}_{\bk_1}{\mathcal
R}_{\bk_2}\mathcal{R}_{\bk_3} \rangle_c$ to the CMB bispectrum is important to include in order to
constrain the deviation from Gaussianity, e.g. in order to test the consistency
relation~\cite{consist}.

\vfill

%%%%%%%%%%%%%%%%%%%%%%%%%%%%%%%%%%%%%%%%%%%%%%%%%%%%%%
\acknowledgements

B.O., P.K.S.D. and C.C. acknowledge support from the NRF (South Africa).

\end{document}